\documentclass[aps,prl,12pt,
nofootinbib]{revtex4-1} 
\usepackage{amsfonts,amsmath,amssymb}   
\usepackage{latexsym}    
\usepackage{bm}          
\usepackage{esvect}      
\usepackage{slashed}     
\usepackage{color}       
\usepackage{verbatim}    
\usepackage{ulem}        

\usepackage{graphicx}    
\usepackage{float}       
\usepackage{subfigure}   

\usepackage[colorlinks=true,
citecolor=magenta,
linkcolor=black,
urlcolor=magenta]{hyperref}

\begin{document}
\title{Toward a Unified Picture of Confinement 
	and Baryon Structure}

\author{Fan Lin}
\email{linfan@aust.edu.cn}
\author{Xinyang Wang}
\email{wangxy@aust.edu.cn}
\affiliation{Center for Fundamental physics, School of Mechanics and Physics, Anhui University of Science and Technology, Huainan, Anhui 232001, China}

  \begin{abstract}
  			In this work, we investigate the infrared structure of quantum chromodynamics from the perspective of the Cho--Faddeev--Niemi decomposition and the Faddeev--Niemi effective theory of Yang--Mills fields. We argue that the topological solitons of the Faddeev--Niemi theory, namely gluon knots characterized by the Hopf invariant, should be regarded as the relevant ground-state degrees of freedom of Yang--Mills theory in the deep infrared region. In this framework, gluon knots provide a unified description of monopole condensation and the center-vortex confinement mechanism. We further propose that baryons are composite objects consisting of quarks immersed in a gluon-knot background. The monopole condensate associated with the gluon knot realizes dual superconductivity, squeezes color-electric flux into flux tubes, and naturally generates the $\mathrm{Y}$-shaped confinement structure of baryons. Simultaneously, the strong local color--magnetic field generated by the gluon knot induces chiral symmetry breaking through magnetic catalysis and topological vacuum fluctuations. We show that the Hopf invariant of the gluon knot is closely related to the topology of Yang--Mills vacua and discuss its connection with the axial anomaly and instanton-induced chiral symmetry breaking. Furthermore, the topological current associated with gluon knots provides a natural carrier of gluon angular momentum and may account for a substantial fraction of the proton spin. The resulting picture establishes a possible connection between the infrared topology of Yang--Mills theory and the internal structure of baryons, providing a unified framework for confinement, chiral symmetry breaking, and baryon structure in QCD.
  			
  	\end{abstract}

\date{\today}

\maketitle

\allowdisplaybreaks{}

\setcounter{table}{0}

 \section{Introduction}
 
\par Baryons account for most of the visible mass in the Universe and are composed of confined quarks together with confined gluons. Their dynamics is governed by quantum chromodynamics (QCD), which exhibits several remarkable nonperturbative phenomena, most notably color confinement and spontaneous chiral symmetry breaking. Phenomenologically, three quarks are tightly confined inside a baryon, and isolated quarks have never been observed. Moreover, the baryon mass is much larger than the sum of the current quark masses, indicating that most of its mass is generated dynamically by QCD. Nevertheless, owing to the intrinsically nonperturbative nature of QCD, the microscopic mechanisms responsible for confinement, chiral symmetry breaking, and the internal structure of baryons remain poorly understood.

\par In the infrared region, where chiral symmetry breaking gives rise to the chiral effective field theory ($\chi$EFT), baryons are successfully described as topological solitons of the effective meson fields: for $N_f\ge2$ they are identified with Skyrmions~\cite{Skyrme:1961vq,Zahed:1986qz,Ma:2016npf}, while for the one-flavor case they appear as vortices on the $\eta'$ domain wall~\cite{Komargodski:2018odf,Ma:2019xtx,Karasik:2020pwu,Karasik:2022tmd,Lin:2023qya,Lin:2025yxh,Lin:2025lzf}. This picture is supported by the large-$N$ limit and anomaly matching arguments~\cite{tHooft:1973alw,tHooft:1979rat}. Since topological properties are generally robust under continuous deformations and can persist across energy scales, even when the effective degrees of freedom change, the topology of these low-energy solitons should ultimately originate from QCD. This strongly suggests that the gluon configuration inside baryons is itself topologically nontrivial.
 
 \par In contrast to quantum electrodynamics, QCD is dominated by the gluon sector, which is described by a non-Abelian $SU(3)$ Yang--Mills theory. Yang--Mills theory is highly nontrivial and is conjectured to exhibit confinement and a mass gap. The gluon fields not only bind quarks inside baryons but also play a central role in inducing chiral symmetry breaking. Therefore, to elucidate the internal structure and dynamics of baryons, it is essential to first identify the relevant ground-state degrees of freedom of Yang--Mills theory in the infrared limit, where the energy scale runs to zero.

 \par For a general non-Abelian $SU(N)$ Yang--Mills theory of gluon fields, ’t~Hooft proposed that the non-Abelian $SU(N)$ theory can be formally described as an Abelian $U(1)^{N-1}$ theory containing both color-electric charges and color-magnetic charges (monopoles)~\cite{tHooft:1981bkw,tHooft2002}. This implies that the diagonal and off-diagonal gluon fields play distinct dynamical roles. The color-magnetic monopoles arise as singular configurations of the off-diagonal gluon fields, with magnetic charges that are topological in nature and classified by the homotopy group $\pi_2\!\left(SU(N)/U(1)^{N-1}\right)\cong\mathbb{Z}^{N-1}$. In order to describe these color-magnetic monopoles, the so-called Cho--Faddeev--Niemi (CFN) decomposition of Yang--Mills theory is proposed~\cite{Cho:1979nv,Faddeev:1998eq,Faddeev:1998yz}. 
 
 In the CFN decomposition, the non-Abelian gauge field is separated into two parts: one describes the color-magnetic monopoles as topological configurations of the off-diagonal gluon fields, while the remaining part forms an Abelian Higgs multiplet composed of diagonal gluon fields and complex scalar fields. It is generally believed that, in the extreme infrared regime, the monopoles condense, so the Abelian Higgs multiplets can be integrated out, leading to an effective action defined on the coset space $SU(N)/U(1)^{N-1}$~\cite{Langmann:1999nn}. This effective action admits a family of topological solitonic solutions in three-dimensional space, classified by $\pi_{3}\!\left(SU(N)/U(1)^{N-1}\right)\cong\mathbb{Z}$, which are commonly referred to as gluon knots. These gluon knots are therefore conjectured to be stable ground-state degrees of freedom of Yang--Mills theory.

 \par If the ground-state degrees of freedom of Yang--Mills theory are indeed gluon knots, it is natural to consider the interaction between quarks and gluon knots in QCD, leading to quark confinement inside baryons. In the dual superconductivity picture, the color-magnetic monopoles originating from the off-diagonal $SU(N)/U(1)^{N-1}$ gluon fields are assumed to condense. This monopole condensate interacts with quarks and squeezes the color-electric flux into flux tubes, which connect quarks and result in confinement. Almost the entire string tension of the flux tube is contributed by the diagonal $U(1)^{N-1}$ gluon fields, a phenomenon known as Abelian dominance. Evidently, the color-magnetic monopole condensate is essential for the formation of baryonic structure. Since a gluon knot represents a specific configuration of the color-magnetic monopole condensate, it is a natural candidate for the monopole condensate inside baryons. In this article, we therefore propose a picture of baryon structure in which a gluon knot exists inside the baryon. Such a gluon knot naturally leads to both confinement and chiral symmetry breaking, in agreement with existing theoretical and phenomenological studies.

\section{Yang-Mills theory and Cho--Faddeev--Niemi decomposition}

  \par In the infrared region, Yang--Mills theory is notoriously difficult to solve directly due to its strong coupling. It has been proposed that the $SU(N)$ Yang--Mills theory flows to the Faddeev--Niemi theory in the infrared region, based on the Abelian projection introduced by 't Hooft~\cite{tHooft:1981bkw,tHooft2002} and the Cho--Faddeev--Niemi decomposition~\cite{Cho:1979nv,Faddeev:1998eq,Faddeev:1998yz,Faddeev:1996zj}. In this section, we briefly review this framework.
  
  \par The Lagrangian of the $SU(N=3)$ Yang--Mills theory is (Throughout this paper, repeated indices are summed over unless otherwise specified.)
  \begin{equation}
  	\mathcal{L}_\mathrm{YM}
  	=-\frac{1}{2g^2}\operatorname{tr}(F_{\mu\nu}F^{\mu\nu})
  	=-\frac{1}{4g^2}(F_{\mu\nu}^a)^2,
  	\qquad
  	F_{\mu\nu}^a
  	=\partial_\mu A_\nu^a-\partial_\nu A_\mu^a+f^{abc}A_\mu^bA_\nu^c,
  \end{equation}
  where the generators of $\mathfrak{su}(N)$ are normalized according to
  \begin{equation}
  	T^aT^b
  	=\frac{1}{2N}\delta^{ab}
  	+\frac{1}{2}d^{abc}T^c
  	+\frac{i}{2}f^{abc}T^c,
  \end{equation}
  with $f^{abc}$ and $d^{abc}$ denoting the antisymmetric and symmetric structure constants, respectively. In a global basis of the Lie algebra, the gauge field $A_\mu$ can be decomposed as
  \begin{equation}
  	A_\mu =  A_{i,\mu} H_i +  A_{\bm{\alpha},\mu} E_{\bm{\alpha}},
  \end{equation}
  where $i=1,\dots,N-1$ labels the Cartan generators $H_i$, and $\bm{\alpha}$ labels the root vectors associated with the ladder operators $E_{\bm{\alpha}}$. This global basis is suitable for describing smooth gauge configurations, but it is not convenient for investigating topological defects associated with internal singularities, such as monopoles, vortices, and knots, all of which will be encountered later. To study these topological defects, especially monopoles, it is advantageous to introduce a local basis for the gauge field. Inspired by the Wu--Yang description of monopoles~\cite{Wu:1976ge}, this leads to the Cho--Faddeev--Niemi decomposition.
  
 \par Monopoles are classified by the homotopy group
 $\pi_2\!\left(SU(N)/U(1)^{N-1}\right)\cong\mathbb{Z}^{N-1}$
 associated with the coset space of the off-diagonal gluons. This implies that monopole configurations can be incorporated through an appropriate choice of gauge. Performing a gauge transformation
 \begin{equation}
 	A_\mu \rightarrow \Omega A_\mu\Omega^{\dagger}+i\Omega(x)\partial_\mu\Omega^{\dagger}, 
 	\label{gauge}
 \end{equation}
 with $\Omega=g(x)$ determined by the monopole configuration, one defines the local Cartan basis
 \begin{equation}
 	m_i=gH_i g^\dagger=m_i^aT^a,
 \end{equation}
 which gives rise to $N-1$ mutually orthogonal unit vectors taking values in the coset space $SU(N)/U(1)^{N-1}$. In terms of this local basis, the gauge field can be decomposed according to the Cho--Faddeev--Niemi (CFN) decomposition~\cite{Cho:1979nv,Faddeev:1998eq,Faddeev:1998yz} 
 \begin{gather}
 	A_\mu^a
 	=C_{i,\mu}m_i^a
 	+f^{abc}\partial_\mu m_i^bm_i^c
 	+\rho_{ij}f^{abc}\partial_\mu m_i^bm_j^c
 	+\sigma_{ij}d^{abc}\partial_\mu m_i^bm_j^c,\\
 	V_\mu^a=C_{i,\mu}m_i^a+H_\mu^a,
 	\qquad
 	H_\mu^a=f^{abc}\partial_\mu m_i^bm_i^c,
 	\qquad
 	X_\mu^a
 	=\rho_{ij}f^{abc}\partial_\mu m_i^bm_j^c
 	+\sigma_{ij}d^{abc}\partial_\mu m_i^bm_j^c.
 \end{gather}
 Here $C_{i,\mu}$ and $X_\mu$ denote the diagonal $U(1)^{N-1}$ gluons and the off-diagonal $SU(N)/U(1)^{N-1}$ gluons, respectively. The fields $\phi_{ij}\equiv\rho_{ij}+i\sigma_{ij}$ form $N(N-1)$ complex scalar fields, which transform under $C_{i,\mu}$ as Abelian--Higgs multiplets. The field $H_\mu$ originates from the off-diagonal part of the pure-gauge term $ig\partial_\mu g^\dagger$ and describes the color-magnetic monopole degrees of freedom. The CFN decomposition leads to the completeness condition 
 \begin{equation}
 	D_\mu^H m_i^{a} = \partial_\mu m_i^a + f^{abc} H_\mu^b m_i^c = 0,
 	\label{completeness}
 \end{equation}
 which implies the field strength of $H_\mu^a$ is totally located in the vector space spanned by $m^a$.

  \par The gluon field strength can be expressed in terms of the CFN decomposition as
  \begin{equation}
  	F_{\mu\nu}^a=F_{\mu\nu}^{V,a}+D_{\mu}^V X_{\nu}^{a}-D_{\nu}^V X_{\mu}^{a}+f^{abc}X_\mu^bX_\nu^c,\quad 
  	D_\mu^V X_\nu^{a}=\partial_\mu X_\nu^a+f^{abc}V_\mu^bX_\nu^c,
  \end{equation}
  where
  \begin{gather}
  	\label{strength}
  	F_{\mu\nu}^{V,a}=\partial_\mu V_\nu^a-\partial_\nu V_\mu^a+f^{abc}V_\mu^b V_\nu^c
  	=m_i^a\!\left(F_{i,\mu\nu}^{C}+F_{i,\mu\nu}^{H}\right),\\
  	F_{i,\mu\nu}^{C}=\partial_\mu C_{i,\nu}-\partial_\nu C_{i,\mu},\quad 
  	F_{i,\mu\nu}^{H}=m_i^a(\partial_\mu H_\nu^a-\partial_\nu H_\mu^a+f^{abc}H_\mu^b H_\nu^c)
  	=f^{bcd}m_i^b\partial_\nu m_j^c\partial_\mu m_j^d .
  \end{gather}
  Accordingly, the $SU(N)$ Yang--Mills Lagrangian becomes
  \begin{equation}
  	\begin{aligned}
  		\mathcal{L}_\mathrm{YM}
  		&= -\frac{1}{4 g^2}  (F_{\mu\nu}^a)^2\\
  		&=-\frac{1}{4 g^2}
  		\Big[
  		(F_{\mu\nu}^{V,a})^2
  		+2f^{abc}F_{\mu\nu}^{V,a} X^{b,\mu}X^{c,\nu}
  		+(D_{\mu}^V X_{\nu}^{a}-D_{\nu}^V X_{\mu}^{a}+f^{abc}X_{\mu}^bX_{\nu}^c)^2
  		\Big].
  	\end{aligned}
  	\label{YM1}
  \end{equation}
 \par The monopoles originate from singular configurations of the off-diagonal gluon fields rather than from fundamental particle excitations~\cite{Cho:1980vs,Cho:1981ze}. In the Cho--Faddeev--Niemi decomposition, the field strength $F_{i,\mu\nu}^{H}$ is associated with the color-magnetic monopole current $k_{i,\nu}$,
 \begin{equation}
 	\partial^\mu \star F_{i,\mu\nu}^{H}=k_{i,\nu}.
 \end{equation}
 So for a closed surface $\Sigma$, the magnetic charge of the $i$-th monopole is~\cite{Cho:1980vs,Cho:1981ze}
 \begin{equation}
 	g_{m,i}= \oint_{\Sigma} F_{i,\mu\nu}^{H}\,\mathrm{d}\sigma^{\mu\nu},
 \end{equation}
 where $\mathrm{d}\sigma^{\mu\nu}$ is the area element.

 \par Isolated color-magnetic charges have not been observed, indicating that the total magnetic charge of the QCD vacuum vanishes. This, however, does not imply that monopoles are unimportant. On the contrary, monopole--antimonopole pairs are continuously created and annihilated and may condense, thereby dominating the Yang--Mills vacuum in the infrared region. Such monopoles themselves must be confined by some additional mechanism. Lattice calculations indicate that the condensate of the off-diagonal gluon fields is stronger than that of the diagonal gluons~\cite{Amemiya:1998jz,Gongyo:2013sha}, suggesting that the off-diagonal sector dominates the gluon condensate. At the classical level, the Yang--Mills Lagrangian $\mathcal{L}_\mathrm{YM}$ contains quartic interaction terms of the form $f^{abc}f^{ade}X_{\nu}^{b}X_{\mu}^{d}X_{\nu}^{e}X_{\mu}^{c}$, which are responsible for generating an effective mass for the off-diagonal gluon fields through condensation~\cite{Kondo:2004dg,Walker:2006jd}. Such massive off-diagonal gluons have been argued to remove instabilities in both the Nielsen--Olesen and Savvidy vacua~\cite{Savvidy:1977as,Nielsen:1978rm,Kondo:2004dg,Walker:2006jd}. The Yang--Mills Lagrangian also contains interaction terms of the form $f^{abc}f^{ade}H_{\nu}^{b}H_{\mu}^{d}X_{\nu}^{e}X_{\mu}^{c}$. As discussed in~\cite{Kondo:2004dg,Walker:2006jd}, these interactions imply that the off-diagonal gluon condensate generates an effective mass for the monopole field $H_\mu^a$, providing a mechanism for monopole confinement. Therefore, understanding the role of topological defects, such as monopoles, is essential for revealing the infrared dynamics of $SU(N)$ Yang--Mills theory.

 \section{monopole, center vortex, gluon knot and confinement}
 
 \par Generally speaking, the condensation of the off-diagonal gluon fields generates a mass gap and confines the monopoles. Consequently, long-range interactions are suppressed, and the color-magnetic flux associated with monopoles is squeezed into vortex tubes (vortex strings). This naturally motivates the search for vortex solutions in the infrared regime of Yang--Mills theory.
 
\par In the framework of the Cho--Faddeev--Niemi decomposition, the Abelianized sector of Yang--Mills theory, which is responsible for the vortex solutions, consists of two parts: the $N-1$ diagonal gluon fields $C_{i,\mu}$ and the $N(N-1)$ off-diagonal gluon fields represented by the complex scalars $\phi_{ij}=\rho_{ij}+i\sigma_{ij}$. These fields together transform as Abelian--Higgs multiplets under the residual $U(1)^{N-1}$ gauge symmetry. Consider a $U(1)^{N-1}$ gauge transformation
\[
\Omega(x)=\exp(i\theta^k m_k).
\]
The local axes $m_i$ remain invariant, while the fields transform as
\begin{equation}
	C_{k,\mu}\rightarrow C_{k,\mu}+\partial_\mu\theta^k,
	\qquad
	\phi_{ij}\rightarrow e^{i\bm{\theta}\cdot\bm{\alpha}^{ij}}\phi_{ij},
\end{equation}
where $\bm{\theta}=(\theta_1,\theta_2,\dots,\theta_{N-1})$ and $\bm{\alpha}^{ij}=(\alpha^{ij}_1,\alpha^{ij}_2,\dots,\alpha^{ij}_{N-1})$ denotes the root vector associated with the $(i,j)$ component. This shows that $\phi_{ij}$ behave as matter fields carrying charges $\bm{\alpha}^{ij}$ under the residual $U(1)^{N-1}$ symmetry.

\par The condensation of the off-diagonal gluons implies that $|\phi_{ij}|^2=\rho_{ij}^2+\sigma_{ij}^2>0$. Since $\phi_{ij}$ indeed form Higgs multiplets under the Abelian gauge fields $\bm{C}_\mu=(C_{1,\mu},C_{2,\mu},\dots,C_{N-1,\mu})$, one can define the covariant derivative
\begin{equation}
	D_\mu^C\phi_{ij}
	=
	\left(
	\partial_\mu
	-i\,\bm{\alpha}^{ij}\cdot\bm{C}_\mu
	\right)\phi_{ij}.
\end{equation}
A nonvanishing vacuum expectation value of $\phi_{ij}$ ensures the existence of vortex solutions~\cite{Mohamadnejad:2013dqa}. We refer to these vortices as CFN vortices, whose fluxes are characterized by the vector
\begin{equation}
	\bm{\Phi}
	=
	(\Phi_1,\Phi_2,\dots,\Phi_{N-1}),
	\qquad
	\Phi_k
	=
	\oint_{\mathcal C}C_{k,\mu}dx^\mu,
\end{equation}
where the integral is taken along a closed loop $\mathcal C$ encircling the vortex. Adopting the ansatz in local cylindrical coordinates,
\begin{equation}
	\phi_{ij}(r,\varphi,z)
	=
	f(r)e^{in\varphi},
	\qquad
	\phi_{ij}(r,\varphi=0,z)
	=
	\phi_{ij}(r,\varphi=2\pi,z),
	\qquad
	n\in\mathbb Z,
\end{equation}
one finds that the allowed flux vector $\bm{\Phi}_{\mathrm{CFN}}$ of a single vortex satisfies the quantization condition
\footnote{We adopt the following conventions for the root and weight vectors. Let $\{\bm e^n\}_{n=1}^N$ be an orthonormal basis of $\mathbb R^N$ satisfying $\bm e^i\cdot\bm e^j=\frac12\delta^{ij}$. The simple roots are $\bm\alpha^n=\bm e^n-\bm e^{n+1}$ ($n=1,\dots,N-1$), such that $|\bm\alpha|^2=1$. The co-root is defined by $\bm\alpha^\vee=2\bm\alpha/|\bm\alpha|^2=2\bm\alpha$. The weights of the fundamental representation are
	$\bm\omega^\ell=\bm e^\ell-\frac1N\sum_{k=1}^N\bm e^k$, with $\ell=1,\dots,N$.}
\begin{equation}
	\bm{\alpha}^{(ij)}
	\cdot
	\bm{\Phi}_{\mathrm{CFN}}
	=
	2\pi n,
	\qquad
	n\in\mathbb Z.
\end{equation}
Therefore, the magnetic fluxes of CFN vortices are quantized and reside on the co-root lattice of the gauge group. In particular, the simplest nontrivial vortex carries
\begin{equation}
	\bm{\Phi}_{\mathrm{CFN}}
	=
	2\pi\bm{\alpha}^{(ij)}.
\end{equation}

\par
\begin{figure}[htbp]
	\centering
	\includegraphics[width=0.6\linewidth]{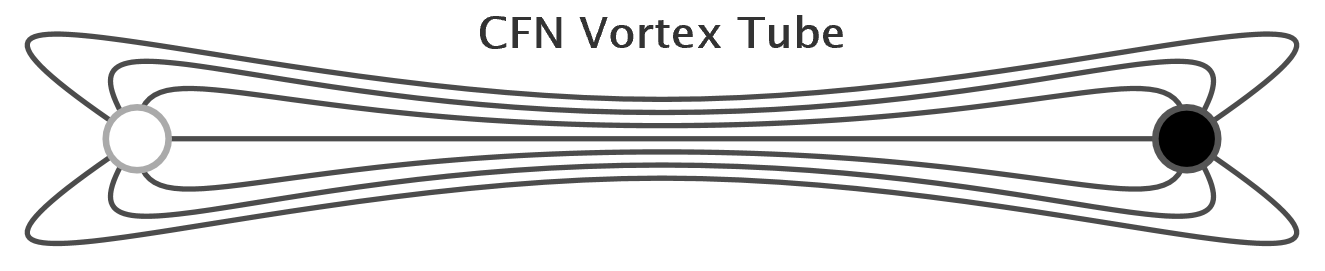}
	\caption{Schematic illustration of a monopole-antimonopole pair confined by a CFN vortex tube.}
	\label{fig:CFN}
\end{figure}

 As is well known, the topological configurations of color-magnetic monopoles in $SU(N)$ Yang--Mills theory are classified by the second homotopy group of the maximal flag manifold,
\begin{equation}
	\pi_2\!\left(SU(N)/U(1)^{N-1}\right)
	\cong
	\mathbb Z^{N-1},
\end{equation}
which implies that the monopole charges are quantized and constrained to the co-root lattice. According to Eq.~(\ref{strength}), the field strengths $F_{i,\mu\nu}^{C}$ associated with the Abelian gauge fields and $F_{i,\mu\nu}^{H}$ arising from the monopole configurations enter on an equal footing. This suggests that the monopole degrees of freedom may behave effectively as charged matter fields and can therefore be confined by the CFN vortices. Consequently, one is naturally led to propose a direct connection between monopoles and CFN vortices: the latter are sourced by color-magnetic monopoles and interpolate between monopole--antimonopole pairs, terminating on them, as illustrated in Fig. \ref{fig:CFN}. However, this picture is not unique, and alternative possibilities remain to be explored.

\par One notes that every root vector $\bm{\alpha}^{ij}$ can be written as the difference of two weights of the fundamental representation, $\bm{\alpha}^{ij}=\bm{\omega}^i-\bm{\omega}^j$, suggesting that a CFN vortex may further fractionate into more elementary constituents. Under the assumption that such a dissociation occurs, the composite adjoint field $\phi_{ij}$ can be factorized into a product of two weight-carrying fundamental scalar fields,
\begin{equation}
	\phi_{ij}=\sigma_i\sigma_j^\dagger.
\end{equation}
Here, $\sigma_i$ and $\sigma_j$ carry the weights $\bm{\omega}^i$ and $\bm{\omega}^j$ of the fundamental representation, respectively, and transform under local $U(1)^{N-1}$ gauge rotations as
\begin{equation}
	\sigma_i\to e^{i\bm{\theta}\cdot\bm{\omega}^i}\sigma_i,
	\qquad
	\sigma_j\to e^{i\bm{\theta}\cdot\bm{\omega}^j}\sigma_j.
\end{equation}
Substituting this factorization into the adjoint covariant derivative $D_\mu^C\phi_{ij}$ gives
\begin{equation}
	\begin{aligned}
		D_\mu^C\phi_{ij}
		&=
		\left[
		\partial_\mu
		-i(\bm{\omega}^i-\bm{\omega}^j)\cdot\mathbf C_\mu
		\right]
		(\sigma_i\sigma_j^\dagger)
		\\
		&=
		\left[
		(\partial_\mu-i\mathbf C_\mu\cdot\bm{\omega}^i)\sigma_i
		\right]
		\sigma_j^\dagger
		+
		\sigma_i
		\left[
		(\partial_\mu-i\mathbf C_\mu\cdot\bm{\omega}^j)\sigma_j
		\right]^\dagger.
	\end{aligned}
\end{equation}
Assuming that both fields acquire nonvanishing vacuum expectation values, individual weight-vortex solutions may emerge. However, the explicit spatial ansatz for $\sigma_i$ and $\sigma_j$ must respect the boundary conditions appropriate to the confinement phase, which are invariant only up to transformations of the center group $Z_N$. Consequently, unlike the regular adjoint field $\phi_{ij}$, the fundamental fields $\sigma_i$ and $\sigma_j$ may acquire a nontrivial $Z_N$ phase factor when transported once around the vortex axis. In local cylindrical coordinates $(r,\varphi,z)$, this twisted boundary condition requires
\begin{equation}
	\sigma_i(r,\varphi=2\pi,z)
	=
	e^{i2\pi n/N}
	\sigma_i(r,\varphi=0,z),
	\qquad
	\sigma_j(r,\varphi=2\pi,z)
	=
	e^{i2\pi n/N}
	\sigma_j(r,\varphi=0,z),
\end{equation}
where $n=1,2,\dots,N-1$. This naturally motivates the fractional winding ansatz
\begin{equation}
	\sigma_i(r,\varphi,z)
	=
	f_i(r)e^{i(n_i-n/N)\varphi},
	\qquad
	\sigma_j(r,\varphi,z)
	=
	f_j(r)e^{i(n_j-n/N)\varphi},
	\qquad
	n_i,n_j\in\mathbb Z.
\end{equation}

\par The allowed magnetic flux vector $\bm{\Phi}_{\sigma_i}$ for an isolated fractional weight-vortex is determined by the corresponding topological quantization condition associated with its phase winding,
\begin{equation}
	\bm{\omega}^i\cdot\bm{\Phi}_{\sigma_i}
	=
	2\pi\left(n_i-\frac{n}{N}\right).
\end{equation}
Although this quantization condition admits an infinite tower of vortex solutions labeled by $n_i\in\mathbb Z$, we restrict our attention to the fundamental $\mathbb Z_N$ sector with $n=1$. The lowest-energy $\sigma$-vortex configuration corresponds to $n_i=0$, which minimizes the magnetic field energy. Using the inner-product relation
\begin{equation}
	\bm{\omega}^i\cdot\bm{\omega}^j
	=
	\frac{1}{2}
	\left(
	\delta^{ij}
	-\frac{1}{N}
	\right),
\end{equation}
the quantization condition can be inverted to obtain the flux vectors carried by the dissociated vortices,
\begin{equation}
	\bm{\Phi}_{\sigma_i}
	=
	4\pi\bm{\omega}^i,
	\qquad
	\bm{\Phi}_{\sigma_j}
	=
	4\pi\bm{\omega}^j.
\end{equation}

\par The flux difference between these two constituent center vortices satisfies
\begin{equation}
	\frac{1}{2}
	\left(
	\bm{\Phi}_{\sigma_i}
	-
	\bm{\Phi}_{\sigma_j}
	\right)
	=
	2\pi
	\left(
	\bm{\omega}^i-\bm{\omega}^j
	\right)
	=
	\bm{\Phi}_{\mathrm{CFN}},
\end{equation}
demonstrating that a macroscopic CFN root vortex may dynamically fractionate into two constituent $\sigma$ vortices carrying weight fluxes. Such a fractional flux structure naturally accommodates monopoles with charges residing on the co-root lattice and provides a topological connection between  $\sigma$ vortices and monopoles. Since both types of vortex configurations may coexist in the QCD vacuum, it is important to determine which configuration is energetically favored.

\par To this end, we consider the kinetic part of the effective Lagrangian for the CFN vortex:
   \begin{equation}
	\mathcal{L}_\mathrm{CFN}=(D\mu^C \phi_{ij}) (D_\mu^C \phi_{ij})^\dagger=\big| \left(\partial_\mu - i \bm{\alpha}^{ij} \cdot\bm{C_{\mu}}\right) \phi_{ij}\big|^2.
	\end{equation}
  The static energy of a vortex tube per unit length $L$ is proportional to the square of its magnetic flux vector $|\bm{\Phi}|^2$. Because a CFN vortex carries the flux of a root vector, $|\bm{\Phi}_\mathrm{CFN}|^2 \propto |\bm{\alpha}^{ij}|^2$. Consequently, the total energy of a macroscopically uniform CFN vortex tube can be estimated as:
  \begin{equation}
  	E_{\mathrm{CFN}} \approx \kappa \cdot L \cdot |\bm{\alpha}^{ij}|^2 = 2\cdot\kappa L,\end{equation}
  	where $\kappa$ is a phenomenological constant representing the characteristic string tension.
\par Expanding $\mathcal{L}_\mathrm{CFN}$ explicitly in terms of the constituent $\sigma$ fields yields:
 \begin{equation}
 	\begin{aligned}
 		\mathcal{L}_\mathrm{CFN}&=\left[(D\mu^{(i)}\sigma_i)\sigma_j^\dagger+\sigma_i(D_\mu^{(j)}\sigma_j)^\dagger\right]\cdot\left[\sigma_j(D^{(i)\mu}\sigma_i)^\dagger+(D^{(j)\mu}\sigma_j)\sigma_i^\dagger\right]\\
 		&=|D_\mu^{(i)}\sigma_i|^2|\sigma_j|^2+|\sigma_i|^2|D_\mu^{(j)}\sigma_j|^2+(D_\mu^{(i)}\sigma_i)\sigma_i^\dagger(D^{(j)\mu}\sigma_j)\sigma_j^\dagger+\sigma_i(D^{(i)\mu}\sigma_i)^\dagger\sigma_j(D_\mu^{(j)}\sigma_j)^\dagger,\end{aligned}
 	\end{equation}
 where the fundamental covariant derivative is defined as $D_\mu^{(i)} \sigma_{i} = \left(\partial_\mu - i \bm{\omega}^{i} \cdot\bm{C_{\mu}}\right) \sigma_{i}$. If the vortices $\sigma_i$ and $\sigma_j$ dissociate and are no longer bound within a tight, singular CFN core, their respective phases $\vartheta_i$ and $\vartheta_j$ fluctuate independently. Consequently, within the functional path integral $\int \mathcal{D}\sigma$, the third and fourth cross-terms vanish identically because they contain rapidly oscillating phase factors proportional to $e^{\pm i(\vartheta_i + \vartheta_j)}$, forcing the non-diagonal correlation functions to vanish, $\langle \sigma_i \sigma_j^\dagger \rangle \to 0$ for $i \neq j$. Under these conditions, the decoupled kinetic Lagrangian governing the $\sigma$ vortices reduces to:
 \begin{equation}
 	\mathcal{L}_\mathrm{\sigma}=|D\mu^{(i)}\sigma_i|^2|\sigma_j|^2+|\sigma_i|^2|D_\mu^{(j)}\sigma_j|^2.
 \end{equation}
 \par Furthermore, assuming the two individual $\sigma$ vortices behave independently, the magnitude of each vortex condensate satisfies $|\sigma_i| \approx |\sigma_j| \approx |\phi_{ij}|$. Since the cross-sectional area of a single dissociated $\sigma$ vortex is approximately half that of a unified CFN vortex core, the individual string tension of a $\sigma$ vortex scales to a factor of $\kappa/2$. Noting that the magnetic flux of these $\sigma$ vortices is anchored to the weight lattice ($\bm{\Phi}_{\sigma_i} \propto 2\bm{\omega}^{i}$ and $\bm{\Phi}_{\sigma_j} \propto 2\bm{\omega}^{j}$), the combined energy of the separated pair of vortex tubes can be parameterized as:
 \begin{equation}
 	E_\sigma \approx \frac{\kappa}{2}\cdot L \cdot (|2\bm{\omega}^{i}|^2 + |2\bm{\omega}^{j}|^2).
 	\end{equation}
 In $SU(N)$ gauge theory, the squared norm of a fundamental weight vector is identically $|\bm{\omega}^{i}|^2=\frac{N-1}{2 N}$. Summing the contributions from both individual tubes, the total energy of the fractionated center vortex system yields:
 \begin{equation}
 	E_\sigma \approx \frac{\kappa}{2} \cdot L \cdot (4|\bm{\omega}^{i}|^2 + 4|\bm{\omega}^{j}|^2)=\frac{2(N-1)}{N}\cdot\kappa L.
 	\end{equation}
 Comparing the two configurations, it is evident that the ratio of their energies satisfies:
 \begin{equation}
 	\frac{E_\sigma}{E_{\mathrm{CFN}}}=\frac{N-1}{N} < 1.
 	\end{equation}
 Specifically for $SU(3)$, this evaluation gives $E_\sigma/E_{\mathrm{CFN}} \simeq 2/3$, meaning that a dissociated pair of $\sigma$ vortices carries only two-thirds of the energy of a intact unified CFN root-vortex. This clear energy minimization indicates that the non-perturbative Yang-Mills vacuum energetically favors the fractionation into independent $\sigma$ vortices.

\par For the scalar fields $\sigma_i$ and $\sigma_j$ carrying the respective color weights $\bm{\omega}^i$ and $\bm{\bm{\omega}}^j$ of the fundamental representation, their minimal isolated configurations ($n=1$) carry magnetic fluxes $\bm{\Phi}_{\sigma_i} = 4\pi\bm{\omega}^i$ and $\bm{\Phi}_{\sigma_j} = 4\pi\bm{\omega}^j$. To establish the non-trivial topological nature of these configurations, we evaluate the expectation value of the $SU(N)$ fundamental Wilson loop along a contour $\mathcal{C}$ enclosing the vortex core:
\begin{equation}
	W(\mathcal{C}) = \frac{1}{N} \sum_{\ell=1}^N \exp\left( \mathrm{i}\bm{\omega}^{\ell} \cdot \oint_\mathcal{C} \bm{C}_{\mu} \mathrm{d}x^\mu \right).
\end{equation}
When the contour $\mathcal{C}$ encircles a single isolated $\sigma_i$ vortex, the circulation of the dual gauge field yields $\oint_\mathcal{C} \bm{C}_\mu \mathrm{d}x^\mu = \bm{\Phi}_{\sigma_i} = 4\pi\bm{\omega}^i$. Substituting this flux quantization condition into the Wilson loop expression triggers the color-weight inner product identity $\bm{\omega}^\ell \cdot \bm{\omega}^i = \frac{1}{2}(\delta^{\ell i} - \frac{1}{N})$, which directly reduces the path integral to a center group phase shift:
\begin{equation}
	W(\mathcal{C}) = \frac{1}{N} \left[ e^{\mathrm{i} 2\pi (1 - 1/N)} + (N-1)e^{-\mathrm{i}2\pi/N} \right] = e^{-\mathrm{i} 2\pi / N}.
	\end{equation}
	Conversely, for a trivial contour enclosing no topological flux, $W(\mathcal{C}) = 1$. This evaluation rigorously confirms that each isolated fractional $\sigma$-vortex behaves strictly as a $\mathbb{Z}_N$ center vortex.
	
	\begin{figure}[htbp]
		\centering
		\includegraphics[width=0.6\linewidth]{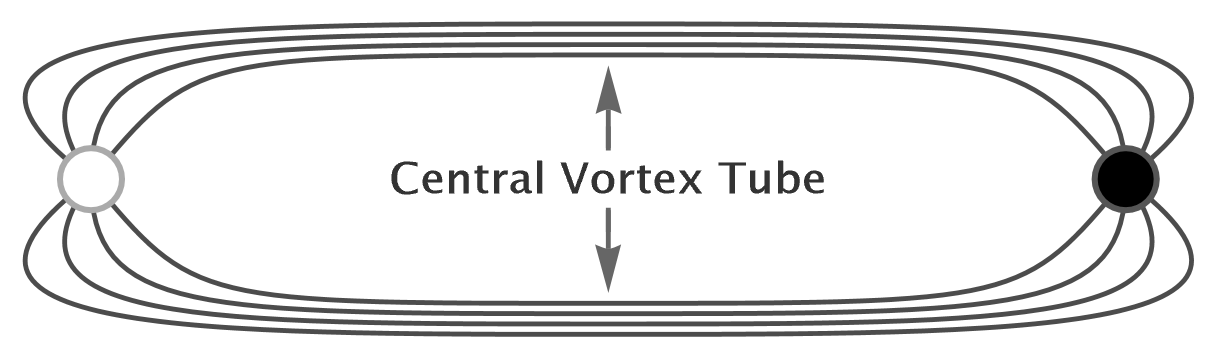}
		\caption{Schematic illustration of a monopole-antimonopole pair confined by a pair of central vortex tubes.}
		\label{fig:CV}
	\end{figure}

	\begin{figure}[htbp]
		\centering
		\includegraphics[width=0.6\linewidth]{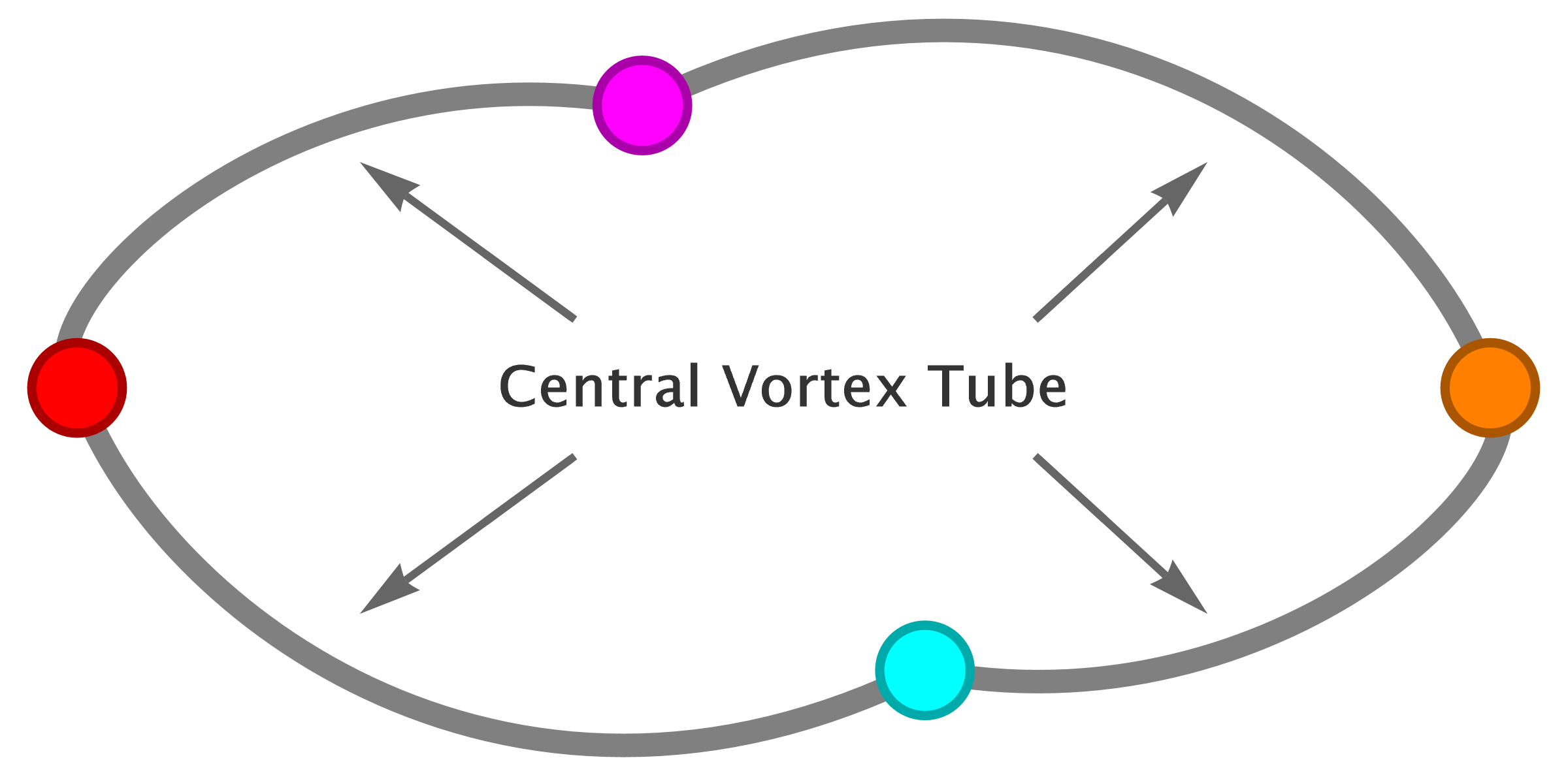}
		\caption{Schematic illustration of four monopoles connected by central vortices to form a long-range loop.}
		\label{fig:Net}
	\end{figure}

	\par \par In this dual-superconductivity confinement scenario, a monopole--antimonopole pair is no longer connected by a single, uniform CFN flux tube. Rather, the root vortex splits into constituent center-vortex tubes shown in Fig.~\ref{fig:CV}, whose long-range closed-loop structure is schematically illustrated in Fig.~\ref{fig:Net}. This picture shares conceptual similarities with the frameworks proposed in Refs.~\cite{Hayashi:2024yjc,Hayashi:2025doq}. Within the non-perturbative Yang-Mills vacuum, a dense ensemble of virtual monopoles undergoes continuous creation and annihilation, accompanied by randomly fluctuating, mutually independent center-vortex pairs (not limited to the basic $\sigma$ vortex we consider above). Consider a macroscopic surface $\mathcal{S}$ bounded by the Wilson loop contour $\mathcal{C}$. Whenever a fluctuating center vortex pierces this surface, the Wilson loop links topologically with the vortex line, forcing the operator to pick up a center element phase shift $z \in \mathbb{Z}_N$, such that $W(\mathcal{C}) \to z W(\mathcal{C})$. Summing over all uncorrelated multi-vortex piercing configurations via a standard Poisson distribution yields the vacuum expectation value:
\begin{equation}
	\langle W(\mathcal{C}) \rangle \sim \exp\left( -\sigma \cdot \text{Area}(\mathcal{S}) \right),
	\end{equation}
	where the effective string tension $\sigma$ is dictated by the spatial density of the independent center-vortex configurations and their corresponding $\mathbb{Z}_N$ phase factors. This explicitly reproduces the area law required for color confinement, validating the center-vortex mechanism directly from the underlying fractionated CFN structure.

  \par \par
\begin{figure}[htbp]
  	\centering
  	\includegraphics[width=0.6\linewidth]{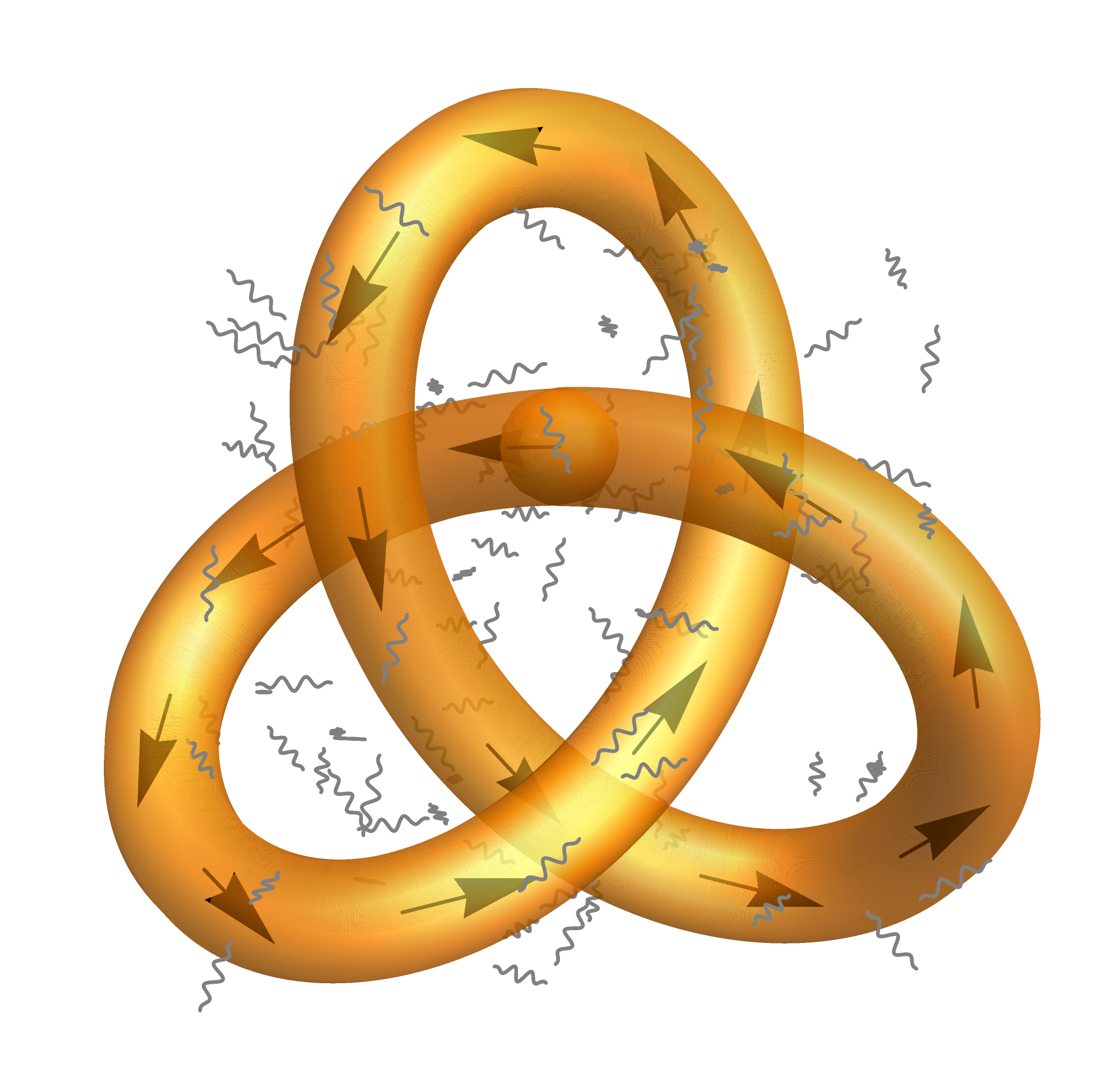}
  	\caption{Schematic illustration of a gluon knot, the simplest Trefoil knot is shown here.  monopoles undergo collective motion, generating color--magnetic currents, with arrows indicating the direction of motion, and the gluon fluctuations are also shown.}
  	\label{fig:knot}
  \end{figure}

 As the energy scale is further lowered toward the deep infrared regime, the separation between monopoles and antimonopoles decreases, and the center-vortex loops formed by their worldlines shrink accordingly. In this limit, a large number of monopoles overlap and effectively form a monopole condensate. Indeed, lattice QCD simulations indicate that off-diagonal gluons acquire a larger effective mass than diagonal gluons~\cite{Amemiya:1998jz,Gongyo:2013sha}. Consequently, the complex scalar fields $\phi_{ij}$, which characterize the off-diagonal gluon degrees of freedom, can be integrated out. The resulting effective theory should then be described by the Abelian gauge fields $C_{i,\mu}$ and the local Cartan basis fields $m_i^a$. However, as shown in Eq.~(\ref{strength}), the field strengths generated by $C_{i,\mu}$ and $m_i^a$ enter on an equal footing and cannot be distinguished in the deep infrared limit. Therefore, the effective theory can be expressed solely in terms of the fields $m_i^a$, making the topological structure of the theory manifest. This effective description of Yang--Mills theory in the deep infrared region is known as the Faddeev--Niemi theory,
  \begin{equation}
  	\mathcal{L}_\mathrm{FN}
  	=(\partial_\mu m_i^a)^2
  	+\frac{1}{e_i^2}(F_{i,\mu\nu}^{H})^2.
  	\label{FN}
  \end{equation}
  Here, $e_i^2$ denote effective couplings and one can easily verify the kinetic term $(\partial_\mu m_i^a)^2=(H_\mu^a)^2$, also behaves as a mass term from the monopole condensate. Owing to the nontrivial homotopy group $\pi_{3}\!\left(SU(N)/U(1)^{N-1}\right)\cong\mathbb{Z}$, this effective Lagrangian admits topologically nontrivial solutions, known as gluon knots \cite{Faddeev:1998eq,Faddeev:1998yz,Faddeev:1996zj}, a schematic illustration of the simplest Trefoil gluon knot shown in Fig. \ref{fig:knot}.
  
  \par Gluon knots are essentially color--magnetic flux tubes that are either self-linked or mutually linked. Their linking structure is classified by the homotopy group $\pi_{3}\!\left(SU(N)/U(1)^{N-1}\right)\cong\mathbb{Z}$ and characterized by the Hopf invariant
  \begin{equation}
  	\mathcal{H}_{(ij)}=\frac{1}{4 \pi^2}\int A_i^H\wedge F_{j}^{H},\qquad \text{with} \qquad \mathrm{d}A_i^H=F_{i}^{H}.
  \end{equation}
  The quantity $\mathcal{H}_{(ij)}$ measures the linking number between the $i$-th and $j$-th color--magnetic flux tubes, while the case $i=j$ corresponds to the self-linking number.  For a gluon knot carrying only the $i$-th color--magnetic flux, it has been shown that its energy is bounded from below by the self-linking number~\cite{vakulenko1979s2,Kundu:1982bc,Faddeev:1996zj,Gladikowski:1996mb},
  \begin{equation}
  	E_i \ge c\, |\mathcal{H}_{(ii)}|^{3/4},	\quad c>0.
  \end{equation}
  This result implies that a gluon knot can carry finite energy and represents a relatively stable configuration of the monopole condensate. Therefore, we propose that gluon knots should be regarded as the fundamental degrees of freedom of Yang--Mills theory in the deep infrared region, unifying the confinement mechanism based on monopole condensation and the center-vortex picture.

We also define the total self-linking number as $\mathcal{H}=\sum_{i}^{N-1}\mathcal{H}_{(ii)}$ for later convenience

\section*{Baryon structure}
\par Now, we take the quarks into account, since gluon knots are regarded as the fundamental degrees of freedom of Yang--Mills theory in the deep infrared region, a natural conjecture is that they can confine quarks and form baryons in QCD. It is well known that quarks are confined within baryons, and different quarks are commonly assumed to be connected by color flux tubes. These flux tubes are generally understood to arise from a background of condensed color–magnetic monopoles. Lattice QCD calculations indicate that almost the entire string tension is contributed by the diagonal gluon fields, a phenomenon known as abelian dominance, which implies that the diagonal components $C_{i,\mu}$ play a central role in quark dynamics. Since gluon knots constitute a particular class of topological color–magnetic monopole condensates, they naturally emerge as promising candidates for the mechanism that confines quark color flux into flux tubes.
\begin{table}[htbp]
	\centering
	\renewcommand{\arraystretch}{1.6}
	\setlength{\tabcolsep}{16pt}
	\begin{tabular}{c|ccc}
		\hline\hline
		& $\psi_{r}$ & $\psi_{g}$ & $\psi_{b}$ \\
		\hline
		$C_{1,\mu}$ charge
		& $-\dfrac{1}{2}$ & $\dfrac{1}{2}$ & $0$ \\
		\hline
		$C_{2,\mu}$ charge
		& $-\dfrac{1}{2\sqrt{3}}$ & $-\dfrac{1}{2\sqrt{3}}$ & $\dfrac{1}{\sqrt{3}}$ \\
		\hline\hline
	\end{tabular}
	\caption{Flux charges of quark fields $\psi$ with colors $r,g,b$ under diagonal $U(1)^2$ gluon fields.}
	\label{charges}
\end{table}

\par From this point, we focus on QCD with $N=3$ colors, which corresponds to the real world. In this case, quark dynamics are dominated by abelian components, and the abelian flux carried by quarks plays the primary role. Under the diagonal $U(1)^2$ gluon fields $C_{1,\mu}$ and $C_{2,\mu}$, quarks of different colors carry different flux charges, as summarized in TABLE~\ref{charges}~\cite{Maedan:1988yi}. Given that gluon knots are stable and robust configurations of monopole condensates, it is natural to adopt a dual magnetic gauge-theory description to analyze their interaction with quarks and the response of the gluon knot to the color flux.

\par A gluon knot is a configuration arising from the condensation of color–magnetic monopoles and can be fully described in terms of complex scalar fields. Quarks interact with gluon knots exclusively through gauge fields, regardless of whether the interaction is electric or magnetic. Motivated by this observation, a dual magnetic gauge-theory Lagrangian was proposed~\cite{Maedan:1988yi,maedan1990abelian,Bardakci:1978ph}:
\begin{equation}
	\begin{split}
		\mathcal{L}_\mathrm{GL}
		= -\frac{1}{4}\,(\vv{M}_{\mu\nu})^{2}
		+ \sum_{\alpha=1}^{3}
		\Bigl[
		\bigl|(\partial_{\mu}+ i g_m \vv{\epsilon}_{\alpha}\!\cdot\!\vv{M}_{\mu})\chi_{\alpha}\bigr|^{2}
		- V\!\left(|\chi_{\alpha}|^{2}\right)
		\Bigr],\\
		\vv{M}_{\mu\nu}
		= \partial_{\mu}\vv{M}_{\nu}-\partial_{\nu}\vv{M}_{\mu}
		+ \epsilon_{\mu\nu\lambda\sigma}\bigl(h^{\lambda}\otimes \vv{j}_{\mathrm{ex}}^{\,\sigma}\bigr),
		\qquad
		h_{\lambda}(x)=-n_{\lambda}(n\!\cdot\!\partial)^{-1}(x).
	\end{split}
\end{equation}
Here, $\vv{M}_{\mu}=(M_{1,\mu},M_{2,\mu})$ are the dual magnetic gauge fields corresponding to the diagonal gluon fields $C_{1,\mu}$ and $C_{2,\mu}$. The coupling $g_m$ is the dual magnetic gauge-theory constant. The vectors
\begin{equation}
	\vv{\epsilon}_1=(1,0),\qquad
	\vv{\epsilon}_2=\left(-\tfrac{1}{2},-\tfrac{\sqrt{3}}{2}\right),\qquad
	\vv{\epsilon}_3=\left(-\tfrac{1}{2},\tfrac{\sqrt{3}}{2}\right)
\end{equation}
form the root lattice on which magnetic charges are distributed. One can verify that $\sum_{\alpha=1}^{3}\vv{\epsilon}_{\alpha}=0$, implying that only two independent types of color–magnetic monopoles exist. The complex scalar fields $\chi_{\alpha}$ describe the condensate density of monopoles with charge $\vv{\epsilon}_{\alpha}$, and a nonzero monopole density is ensured by the Higgs-type potential $V(|\chi_{\alpha}|^{2})$. This potential is suggested to emerge from summing over all monopole trajectories~\cite{Stone:1978mx,Samuel:1979mq,Unsal:2008ch}. The vector $n_{\lambda}$ is an arbitrary constant four-vector, and $\vv{j}_{\mathrm{ex}}^{\,\sigma}$ denotes an external charged current. For instance, if a quark with charge $\vv{Q}$ is at $\bm{a}$ and an antiquark with charge $-\vv{Q}$ is at $\bm{b}$, the external current is
\begin{equation}
	\vv{j}_{\mathrm{ex}}^{\,\mu}
	=\vv{Q}\, g^{\mu 0} \bigl\{\delta(\bm{x}-\bm{a})-\delta(\bm{x}-\bm{b})\bigr\},
	\label{ecurrent}
\end{equation}
with $\vv{Q}=g\,(0,1/\sqrt{3})$ for the $\bar{\psi}_{b}\psi_{b}$ configuration in TABLE~\ref{charges}. Accordingly, the term
$\epsilon_{\mu\nu\lambda\sigma}\bigl(h^{\lambda}\otimes \vv{j}_{\mathrm{ex}}^{\,\sigma}\bigr)$
provides a natural mechanism to incorporate quark dynamics in the presence of a gluon knot~\cite{Zwanziger:1970hk}. 

\par Clearly, the dual magnetic gauge theory $\mathcal{L}_{\mathrm{GL}}$ is of the Ginzburg–Landau type, demonstrating that the interaction between quarks and gluon knots can be consistently described within the framework of dual superconductivity.

\par Since quark dynamics are abelian dominated, it is reasonable to consider only the diagonal color–electric fields generated by quarks. If only three quarks, $\psi_r$, $\psi_g$, and $\psi_b$, are present in space and their diagonal color–electric fluxes have not yet formed flux tubes, they still attract each other via a Coulomb potential, because their charges satisfy
\begin{equation}
	\vv{Q}_{\psi_r}+\vv{Q}_{\psi_g}+\vv{Q}_{\psi_b}=0,
	\label{charge}
\end{equation}
as listed in Table~\ref{charges}. However, the presence of gluon knots alters this scenario. Gluon knots are topological configurations of the monopole condensate, giving rise to nonzero vacuum expectation values of the complex scalar fields $\chi_{\alpha}$. These nonzero values spontaneously break the magnetic gauge symmetry $U(1)^2_m$ (with the subscript $_m$ denoting ``magnetic'' symmetry), causing both the scalar fields $\chi_{\alpha}$ and the dual gauge fields $\vv{M}_{\mu}$ to acquire masses, $m_\chi$ and $m_M$, respectively.

\par It is noteworthy that lattice calculations indicate the diagonal gluon fields are lighter than the off-diagonal ones~\cite{Amemiya:1998jz,Gongyo:2013sha}, implying $m_M < m_\chi$ and suggesting that the QCD vacuum behaves as a weak type-II dual superconductor, consistent with lattice QCD studies~\cite{DAlessandro:2005omg,Cardoso:2010kw}.Quarks immersed in the gluon-knot background radiate diagonal color fields that extend through space. Once these diagonal color fields interact with the gluon knots, they acquire mass and are inhibited from freely penetrating the knots, realizing a dual Meissner effect. Nevertheless, because the monopole condensate is intrinsically inhomogeneous—with higher density along the axis of a gluon knot than in the surrounding regions—the diagonal color fields can partially penetrate through areas of weaker monopole condensation, forming color flux tubes analogous to magnetic flux tubes in type-II superconductors.

\par
\begin{figure}[htbp]
	\centering
	\includegraphics[width=0.6\linewidth]{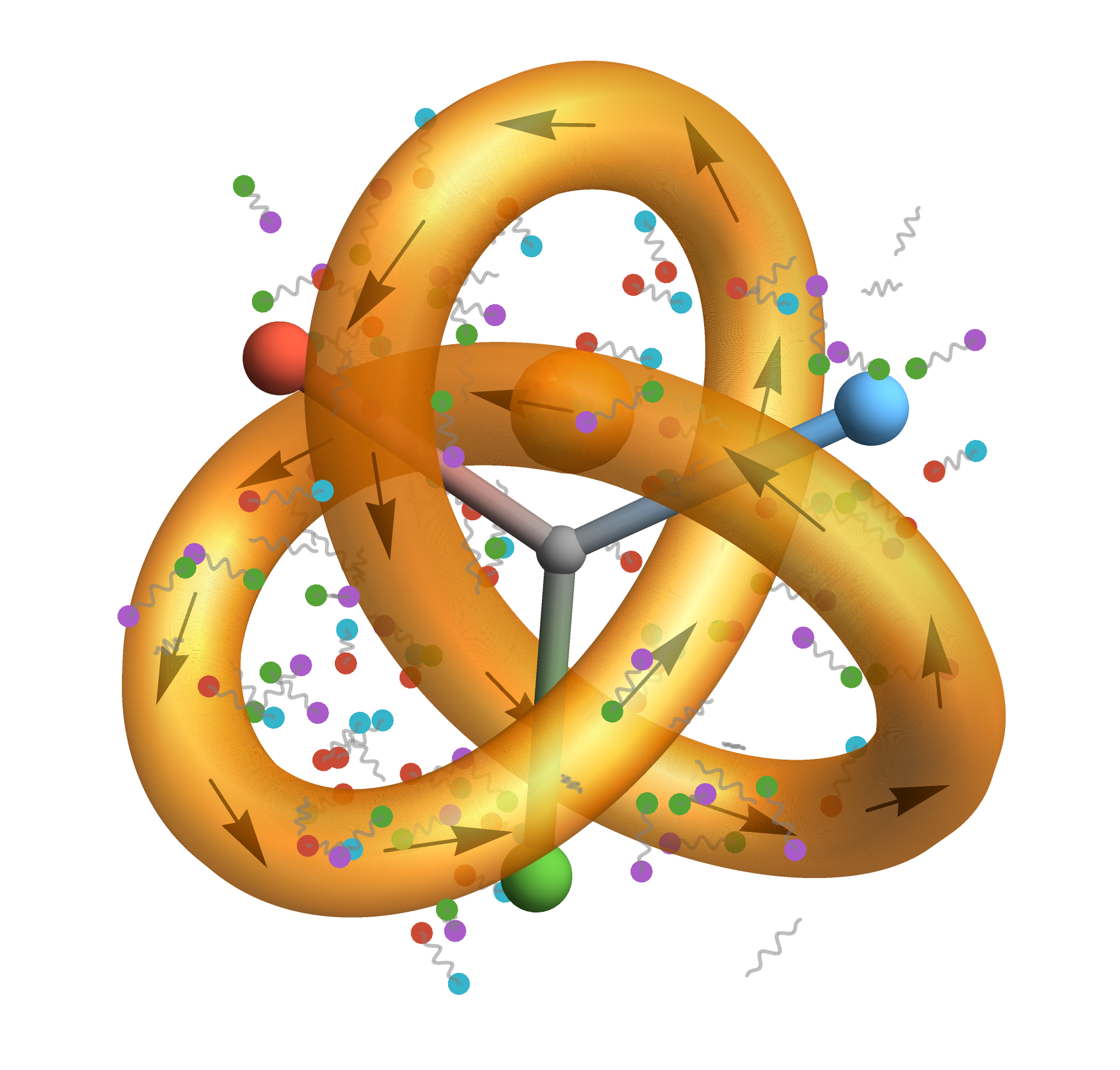}
	\caption{Schematic illustration of the conjectured baryon structure. Three valence quarks are immersed in and confined by the gluon-knot background, which forms the dense core of the baryon.}
	\label{fig:baryon}
\end{figure}

 Although the diagonal color fields of the quarks extend in all directions, the presence of a gluon knot endows them with mass. The mass term for the color fields breaks the Gauss law for flux, which would conflict with the Dirac quantization condition. To resolve this, the color flux of the quarks is forced to be squeezed into flux tubes~\cite{Guimaraes:2012tx}. These flux tubes extend through regions of weaker monopole condensation and eventually connect with each other. Since the QCD vacuum behaves as a weak type-II superconductor~\cite{DAlessandro:2005omg,Cardoso:2010kw}, the flux tubes tend to maximize contact with the gluon knot, lowering the interface energy. As a result, the color flux tubes emitted by the three quarks inside a baryon repel each other, leading to a $\mathrm{Y}$-shaped junction~\cite{Ichie:2002dy}. The gluon knot wraps around the flux tubes, forming the conjectured baryon structure shown in Fig.~\ref{fig:baryon}, with virtual particle fluctuations in the QCD vacuum also depicted. 

\par It should be emphasized that the gluon knot inside a baryon is a continuously evolving dynamical object. Its topological charge is not conserved, and the associated linking number may vary with time. Nevertheless, configurations corresponding to the lowest allowed energy are expected to be the most probable. Even for a fixed topological number, the equations of motion derived from the Lagrangian $\mathcal{L}_{\mathrm{FN}}$ do not uniquely determine a gluon-knot solution unless a specific ansatz is imposed~\cite{Faddeev:1996zj}. Similarly, a gluon knot can wrap the color flux tube with different winding numbers at different times. For a given wrapping configuration, the internal color–magnetic currents of the gluon knot generate a squeezing force that confines the quark color flux into a flux tube. As the positions of the quark charges change over time, the shape and length of the color flux tubes adjust accordingly. As the quark positions evolve, the geometry and length of the flux tubes adjust dynamically. After averaging over all admissible configurations of the gluon knot, color flux tubes, and virtual particle fluctuations, the resulting expectation value exhibits a avenge $\mathrm{Y}$-shaped junction that carries baryon number~\cite{PhysRevD.101.065011}, consistent with recent experimental evidence~\cite{STAR:2024lvy}.

\par When a gluon knot wraps a color flux tube, the dual magnetic gauge field $\vv{M}_{\mu\nu}$ generates a color–magnetic current density 
\begin{equation}
	\vv{J}^{\,\mathrm{m}}_{\nu} = \partial_{\mu} \vv{M}^{\mu\nu},
\end{equation}
which circulates around the axis of the flux tube. If the dominant color–magnetic current is induced by the curl of the color–electric field, $\vv{\bm{J}}^{\,\mathrm{m}} = \bm{\nabla} \times \vv{\bm{E}}$, one obtains a confining force density 
\begin{equation}
	\vv{\bm{F}} = \vv{\bm{J}}^{\,\mathrm{m}} \times \vv{\bm{E}},
\end{equation}
directed toward the flux-tube axis. This ``Maxwell picture'' of confinement is supported by recent studies~\cite{Cea:2023ioo,Baker:2024peg}. 

\par Once a color flux tube is formed, quark confinement is realized. For an almost static quark carrying charge $\vv{Q}$ and its antiquark separated by a distance $r$, the external current has the form in Eq.~(\ref{ecurrent}), and the effective interaction potential takes the Cornell form~\cite{Maedan:1988yi,maedan1990abelian}:
\begin{equation}
	V(r) \simeq -\frac{\vv{Q}^{\,2}}{4\pi}\,\frac{e^{-m_M r}}{r} 
	+ \frac{\vv{Q}^{\,2} m_M^{2}}{4\pi}\, K_0\!\left(\frac{\sqrt{2}\, m_M}{m_\chi}\right) r ,
\end{equation}
where $K_0(x)$ is the modified Bessel function of the second kind. This indicates that a quark–antiquark pair is confined, consistent with results in Refs.~\cite{Suganuma:1993ps,Akhmedov:1995mw}. For three quarks inside a baryon, each quark interacts simultaneously with the other two through the color flux tubes. Owing to the charge relation in Eq.~(\ref{charge}), this interaction is effectively analogous to that between a quark and an antiquark, ensuring confinement of all three quarks inside the baryon.

\par A gluon knot is constituted by a condensate of color–magnetic monopoles. The corresponding gauge-field component
\[
H_\mu \equiv H_\mu^a T^a = f^{abc}\, \partial_\mu m_i^b\, m_i^c\, T^a
\]
generates a locally sizable and approximately uniform color–magnetic field. One can then consider the direct coupling of quark fields $\psi$ to $H_\mu$:
\begin{equation}
	\mathcal{L} = \bar{\psi}\, (i \slashed{\partial} - \slashed{H})\, \psi.
\end{equation}
It is well known that a locally uniform magnetic field induces the magnetic catalysis effect for Dirac fermions. For the quark field $\psi_b$ carrying color charge $\vv{Q} = g\,(0,1/\sqrt{3})$, magnetic catalysis leads to a nonzero chiral condensate~\cite{Gusynin:1994re,Gusynin:1994xp,Gusynin:1998zq,Shovkovy:2012zn}:
\begin{equation}
	\langle 0 | \bar{\psi} \psi | 0 \rangle \sim |g\, \bm{B}_2|, \label{chiral1}
\end{equation}
where $B_2^i = \tfrac{1}{2} \epsilon^{0ijk} F^{H}_{2,ij}$ denotes the locally significant uniform color–magnetic field. Therefore, the presence of a gluon knot induces a chiral condensate of quarks and contributes to the baryon mass.

\par  Non-zero chiral condensate leads to spontaneous chiral symmetry breaking and Nambu--Goldstone boson fields are generated. Consequently, the baryon is immersed in a background of Nambu--Goldstone boson fields, such as the pion field. In the infrared regime, baryon interactions are thus mediated predominantly by the exchange of relatively light mesons, as first anticipated by Yukawa. In this sense, the proposed picture of baryon structure naturally explains chiral symmetry breaking.

\par For the proton, whose mass $M_P$ can be decomposed via the trace anomaly~\cite{Ji:1994av},
\begin{equation}
	2 M_P^{2} = \langle P | \frac{\beta(g)}{2g} F^2 | P \rangle 
	+ \langle P | (1+\gamma_m)\, \bar{\psi} m \psi | P \rangle,
\end{equation}
our picture implies that the gluon knot and the associated flux tubes contribute directly to the dominant part of the gluon condensate, while gluon knots and instantons indirectly generate most of the chiral condensate. According to lattice QCD results~\cite{Yang:2018nqn}, the gluon condensate contributes about $40\%$ of the proton mass, leading to an estimate of the gluon knot inside the proton of roughly $400 \pm 100~\mathrm{MeV}$. The gluon knot naturally forms a relatively dense core of the baryon, whereas quarks are distributed in more peripheral regions. A direct corollary is that the proton charge radius is larger than its mass radius, in agreement with experimental measurements~\cite{RevModPhys.94.015002,PhysRevD.104.054015}.

\par In this picture, both confinement and chiral symmetry breaking emerge as dynamical consequences of gluon knots, with the chiral condensate predominantly distributed in the outer regions of the baryon. As the temperature increases, the chiral condensate is expected to melt before deconfinement occurs. Consequently, the chiral symmetry restoration temperature should be lower than the confinement/deconfinement temperature, consistent with the widely accepted understanding.

\section{baryon structure and axial anomaly}

  \par In general, the number of independent topological invariants of a theory is much smaller than the number of local dynamical degrees of freedom. In particular, there exists the natural isomorphism
  \begin{equation}
  	\pi_{3}\!\left(SU(N)/U(1)^{N-1}\right)
  	\cong
  	\pi_{3}\!\left(SU(N)\right)
  	\cong
  	\mathbb{Z}.
  	\label{isomorphism}
  \end{equation}
  Since $\pi_{3}(SU(N))$ characterizes the topology of Yang--Mills vacua, this suggests that gluon knots may be closely related to the topological structure of the vacuum. The homotopy group $\pi_{3}(SU(N))$ is characterized by the Chern--Simons number of the gauge field $A_\mu$,
  \begin{equation}
  	N=\int d^3\bm{x}\,K^0,
  	\qquad
  	K^\mu
  	=
  	\frac{1}{8\pi^2}
  	\epsilon^{\mu\nu\rho\sigma}
  	\mathrm{tr}
  	\left(
  	A_\nu\partial_\rho A_\sigma
  	-\frac{2i}{3}A_\nu A_\rho A_\sigma
  	\right).
  \end{equation}
  All finite-action field configurations asymptotically approach a pure gauge configuration,
  \begin{equation}
  	x\rightarrow\infty:
  	\quad
  	A_\mu
  	\rightarrow
  	ig(x)\partial_\mu g^\dagger(x).
  \end{equation}
  Within the framework of the CFN decomposition, the topological information is encoded in the gauge transformation of Eq.~(\ref{gauge}) with $\Omega=g(x)$. The corresponding winding number of the vacuum is therefore
  \begin{equation}
  	N(g)
  	=
  	\frac{1}{24\pi^2}
  	\int_{\mathbb{S}_\infty^3}
  	d^3S\,
  	\epsilon^{ijk}
  	\mathrm{tr}
  	\left(
  	g\partial_i g^\dagger
  	\,
  	g\partial_j g^\dagger
  	\,
  	g\partial_k g^\dagger
  	\right),
  \end{equation}
  where the spatial manifold $\mathbb{R}^3$ has been compactified to the three-sphere $\mathbb{S}^3_\infty$.
  
  \par If gluon-knot configurations are present in the vacuum, the corresponding Hopf linking number is nonzero. However, one should not naively identify the Hopf invariant with the vacuum winding number, since the isomorphism (\ref{isomorphism}) does not necessarily imply that the corresponding topological invariants are identical. Using differential forms, the total Hopf self-linking number can be written in terms of the Kirillov two-form as
  \begin{equation}
  	\mathcal{H}
  	=
  	\frac{1}{4\pi^2}
  	\int
  	A_i^H\wedge F_i^H
  	=
  	\frac{1}{\pi^2}
  	\int
  	\mathrm{tr}(H_i\,gdg^\dagger)
  	\wedge
  	\mathrm{tr}
  	\left(
  	H_i\,gdg^\dagger\wedge gdg^\dagger
  	\right).
  	\label{topocharge}
  \end{equation}
  Making use of the completeness relation in Eq.~(\ref{completeness}), which implies that $F_i^H$ spans the Cartan subspace in color space, one finds that the vacuum winding number is related to the total Hopf self-linking number by
  \begin{equation}
  	N(g)=\frac{\mathcal{H}}{4}.
  \end{equation}
  This relation suggests that the emergence of gluon-knot configurations can result in a nontrivial topological vacuum.
  
   \par For QCD in the real world, the number of colors is $N=3$, implying the existence of two types of gluon knots with self-linking numbers $\mathcal{H}_{(ii)}$, $i=1,2$, associated with the two Cartan generators of $SU(3)$. Both types are expected to be present inside baryons and participate in quark confinement. If the vacuum surrounding the baryon tends to be topologically trivial, one may conclude that the self-linking numbers of the two types of gluon knots should have opposite signs, such that the total self-linking number satisfies
   \begin{equation}
   	\mathcal{H}
   	=
   	\mathcal{H}_{(11)}
   	+
   	\mathcal{H}_{(22)}
   	=
   	0.
   \end{equation}
   However, instanton--anti-instanton fluctuations (or other processes carrying a nontrivial second Chern class) may exist in the vacuum. Consequently, baryons may also possess nonzero total self-linking and reside in a topologically nontrivial vacuum sector.
   
   \par As is well known, at the quantum level the axial current
   $
   j^{5,\mu}
   =
   \bar{\psi}\gamma^\mu\gamma^5\psi
   $
   suffers from the axial anomaly~\cite{Adler:1969gk,Bell:1969ts},
   \begin{equation}
   	\partial_\mu j^{5,\mu}
   	=
   	\frac{g^2N_f}{8\pi^2}
   	\mathrm{tr}
   	(F_{\mu\nu}
   	\star F^{\mu\nu}),
   	\quad
   	\star F^{\mu\nu}
   	=
   	\frac{1}{2}
   	\epsilon^{\mu\nu\rho\sigma}
   	F_{\rho\sigma}.
   \end{equation}
   A change in the total self-linking number implies that the topological vacuum must tunnel through instanton configurations, resulting in the emergence of fermionic zero modes and subsequently inducing chiral symmetry breaking through the effective 't~Hooft vertex~\cite{tHooft:1976rip,tHooft:1976snw,Diakonov:2002fq}. The chiral condensate generated either by magnetic catalysis in Eq.~(\ref{chiral1}) or by instantons through the mechanism discussed here can ultimately be understood as originating from a nonzero mean density of zero modes $\overline{\nu(0)}$, according to the Banks--Casher relation~\cite{Banks:1979yr,Diakonov:1984vw,Diakonov:1995ea},
   \begin{equation}
   	\langle\bar{\psi}\psi\rangle
   	=
   	-\frac{\pi}{V}
   	\mathrm{sign}(m)
   	\overline{\nu(0)},
   \end{equation}
   where $m$ denotes the bare quark mass.

  \par The existence of gluon knots inside baryons may also provide further insight into the baryon spin structure. It is well known that quarks carry only a small fraction of the baryon spin because the axial current is not conserved at the quantum level~\cite{EuropeanMuon:1987isl,EuropeanMuon:1989yki}. Although the quark spin can be transferred to gluon spin (or helicity), the mechanism by which gluon fields carry spin remains unclear. So far, we have mainly analyzed the topological charge, while the corresponding current associated with the gluon knot has been largely neglected.
  
  Since the Lagrangian $\mathcal{L}_{\mathrm{FN}}$ does not uniquely determine a gluon-knot solution unless a specific Ansatz is imposed~\cite{Faddeev:1996zj}, even configurations with vanishing total self-linking number in Eq.~(\ref{topocharge}) generally possess nontrivial local topological charge distributions. These local structures naturally generate the spatial current
  \begin{equation}
  	\bm{H}
  	=
  	\frac{1}{8\pi^2}
  	\bm{A}_{i}^{H}
  	\times
  	\bm{E}_{i}^{H},
  \end{equation}
  where bold symbols denote spatial vectors and $\bm{E}_{i}^{H}$ represents the color-electric component of $F_i^H$. This contribution enters the Jaffe--Manohar proton spin decomposition and contributes to the gluon spin component~\cite{Jaffe:1989jz}. Assuming Abelian dominance inside baryons, one may further infer that $\bm{H}$ provides the dominant contribution to the gluon-spin fraction of the proton spin.
  
  Furthermore, the color-electric field $\bm{E}_{i}^{H}$ is induced by the time variation of the color-magnetic field $\bm{B}_{i}^{H}$, and the two fields are locally perpendicular to each other. Consequently, they contribute to the Ji decomposition of the proton spin through the angular momentum density
  \begin{equation}
  	\bm{x}
  	\times
  	\left(
  	\bm{E}_{i}^{H}
  	\times
  	\bm{B}_{i}^{H}
  	\right),
  \end{equation}
  which corresponds to the gluon angular momentum contribution~\cite{Ji:1996ek,Ji:2020ena}. Therefore, one may conclude that a substantial part of the gluon-spin contribution to the proton spin is carried by gluon knots.

\section*{Discussion and Conclusion}
   \par
\begin{figure}[htbp]
   	\centering
   	\includegraphics[width=0.6\linewidth]{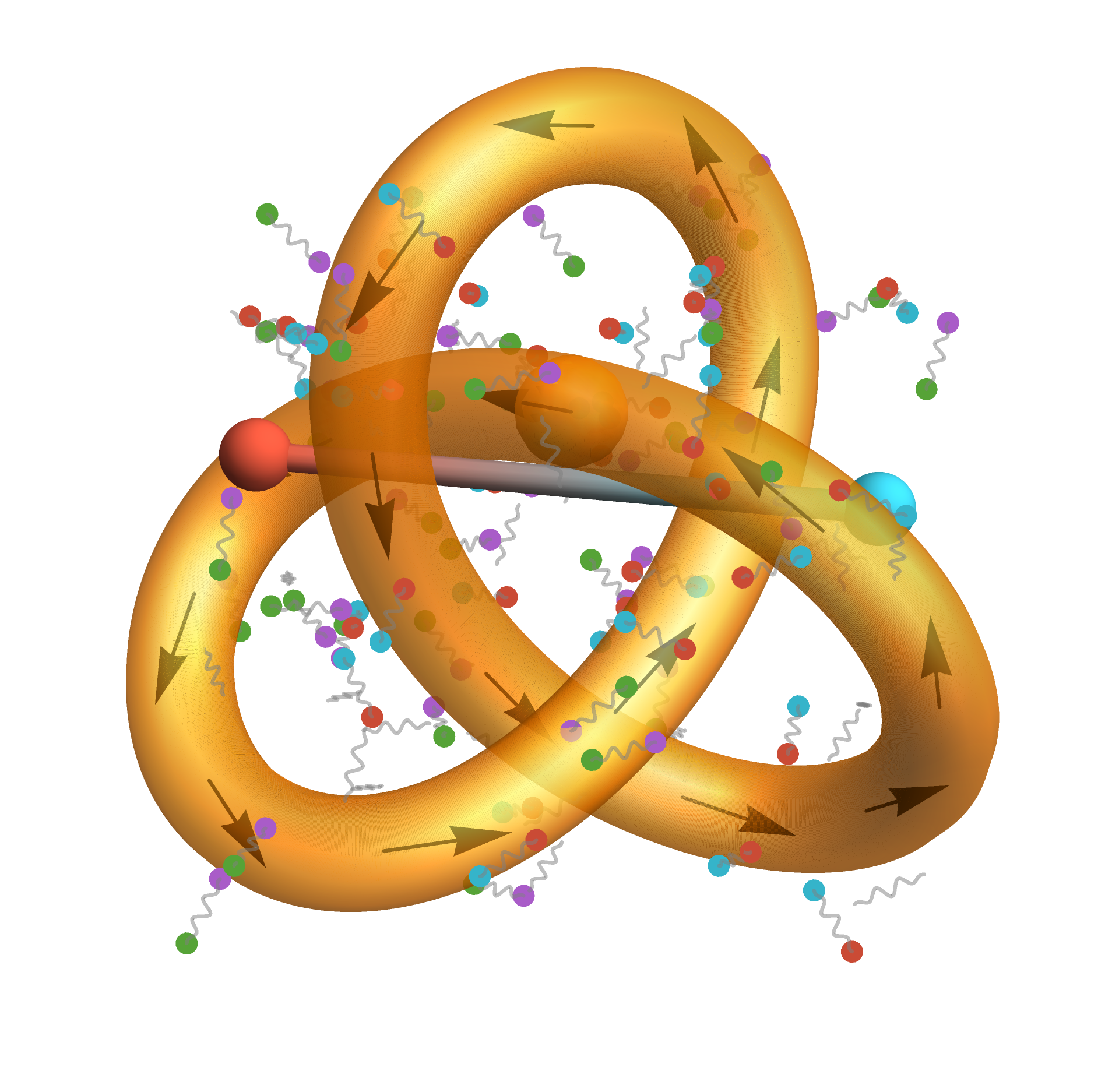}
   	\caption{Schematic illustration of the conjectured heavy-flavor meson structure. A heavy-flavor quark--antiquark pair is immersed in and confined by the gluon-knot background, forming the core of the meson.} 
   	\label{fig:meson}
   \end{figure}

   In the discussion above, we have proposed that baryons possess a gluon-knot structure. The situation for mesons, however, is different. For mesons composed of heavy quarks, such as $J/\psi$, the quark and antiquark are non-relativistic, allowing the formation of a stable flux tube that produces confinement. We therefore conjecture that heavy-flavor mesons may also contain a gluon knot, analogous to the baryon, as illustrated in Fig.~\ref{fig:meson}.
   
   \par
   By contrast, for light-flavor mesons, such as the pseudoscalar mesons $\pi$ and $K$, or the vector mesons $\rho$ and $\omega$, the flux tube is continuously broken by quark--antiquark pair creation. Hence, a different confinement mechanism must operate. A plausible candidate is the Gribov--Zwanziger confinement scenario~\cite{Gribov:1977wm,Zwanziger:1991gz,Zwanziger:1998ez}, which restricts gauge fields to the Gribov region, suppresses infrared gluons, and produces a propagator
   \[
   D(k^2) = \frac{k^2}{k^4 + \lambda^4},
   \]
   naturally leading to quark confinement and dynamical chiral symmetry breaking.
   
   \par
   However, the absence of a persistent color flux tube in light-flavor mesons does not imply the absence of gluon knots. Gluon knots decay only via strong interactions, suggesting that all mesons whose decays are dominated by strong interactions may contain a gluon-knot component, such as vector mesons $\rho,\omega$. Nambu--Goldstone bosons, such as $\pi$ and $K$, are too light to contain gluon knots.
   
   \par
   A particularly interesting case is the $f_0(500)$ resonance, or sigma meson, whose mass is close to the estimated gluon-knot mass inside the proton, $400 \pm 100~\mathrm{MeV}$. We therefore conjecture that $f_0(500)$ may be primarily a gluon knot with an admixture of quark--antiquark components. Similarly, the axial-singlet meson $\eta'$ may also contain a gluon-knot component, consistent with its unusually large mass. More boldly, one may speculate that all glueballs are gluon knots with distinct topological configurations. 
   
   \par
   In conclusion, we propose that the gluon knot represents a fundamental dynamical degree of freedom of Yang--Mills theory. It is natural to conjecture that baryons contain gluon knots, which gives rise to both confinement and chiral symmetry breaking. This picture is consistent with existing theoretical and experimental evidence, although further investigations are required to fully validate and explore its implications.

\section*{Acknowledgements}
This work was supported by the National Natural Science Foundation of China under Grant No. 12547138 and the Scientific Research Foundation for Introduced Talents of Anhui University of Science and Technology.

\bibliographystyle{unsrt}   
\bibliography{refs}


\end{document}